\documentclass[journal]{IEEEtran}
\usepackage{cite}

\ifCLASSINFOpdf
\else
\fi
\usepackage{epsfig}

\usepackage{array}

\usepackage{mdwmath}
\usepackage{mdwtab}

\usepackage[tight,footnotesize]{subfigure}

\usepackage[caption=false]{caption}
\usepackage[font=footnotesize]{subfig}
\usepackage{fixltx2e}
\usepackage{url}

\begin{document}
%
\title{Use of a $d$-Constraint During LDPC Decoding in a Bliss Scheme}
%
%
%

\author{Andries~P.~Hekstra,~\IEEEmembership{Member,~IEEE
}
\thanks{Andries Hekstra is with NXP Semiconductors, High Tech Campus 32,
5656 AE Eindhoven, The Netherlands. This work was done while at Philips Research. Email :
andries.hekstra@nxp.com}
}
\maketitle

\begin{abstract}
Bliss schemes of a run length limited (RLL) codec in combination with an LDPC codec, generate LDPC parity bits over a systematic sequence of RLL channel bits that are inherently redundant as they satisfy e.g. a $d=1$ minimum run length constraint. That is the subsequences consisting of runs of length $d=1$, viz. $\ldots010\ldots$ and $\ldots101\ldots$, cannot occur. We propose to use this redundancy during LDPC decoding in a Bliss scheme by introducing additional $d$-constraint nodes in the factor graph used by the LDPC decoder. The messages sent from these new nodes to the variable or codeword bit nodes exert a ``force'' on the resulting soft-bit vector coming out of the LDPC decoding that give it a tendency to comply with the $d$-constraints. This way, we can significantly reduce the probability of decoding error. 
\end{abstract}

\begin{IEEEkeywords}
Bliss schemes, RLL codes, LDPC codes, factor graph, modified concatenation.
\end{IEEEkeywords}

%
\IEEEpeerreviewmaketitle

\section{Introduction}
%
%
%
%

\newcommand{\mM}{{\cal M}}
\newcommand{\mN}{{\cal N}}
\newcommand{\VAR}{\mbox{VAR}}
\newcommand{\CHK}{\mbox{CHK}}
\newcommand{\CO}{\mbox{CO}}
\newcommand{\sign}{\mbox{signBit}}
\newcommand{\RLDPC}{R_{LDPC}}
\newcommand{\RRLL}{R_{RLL}}

\begin{figure}[H]
\centering
\includegraphics[viewport=0 0 600 400, width=0.95\columnwidth, clip]{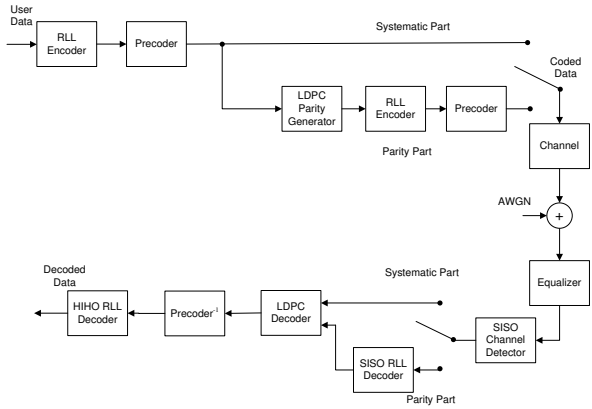}
\caption{Block diagram of a Bliss scheme. 
\label{fig:bliss}}
\end{figure}

\IEEEPARstart{B}{liss} schemes \cite{Bliss}, also called \textit{modified concatenation} schemes by Fan \cite{Fan}, place a modulation encoder and decoder as outer parenthesis and a systematic ECC encoder and decoder as inner parenthesis around the storage channel. This way, the modulation decoder follows the ECC decoder, rather than precedes it, thus, avoiding error propagation by the modulation decoder at the input of the ECC decoder as in the \textit{standard concatenation} scheme \cite{Fan}. As shown in Fig.~\ref{fig:bliss}, the parity bits generated from the sequence of modulation encoded channel bits are encoded with a second modulation encoder, where both modulation encoders are of the \textit{run length limited} (RLL) type \cite{Immink} . 

The so called \textit{$1T$-precoder} in Fig.~\ref{fig:bliss} performs an integration modulo-2 in order to convert \textit{differential bits}, also called \textit{$d,k$-bits} within the context of RLL coding, to so called \textit{unipolar bits,} that are indicative of the type of run on the storage medium. The channel SISO detector is shared between the systematic and the parity part of the LDPC codeword and produces soft-decision values for the unipolar bits. The equalizer filter shortens the channel impulse response. The SISO RLL decoder operates on these soft-decision unipolar bits produced by the channel SISO detectors. Hence, the number of states of the RLL SISO decoder is twice the number expected if differential bits were used, as the state incorporates an additional \textit{polarity bit}\footnote{
The use of a unipolar RLL SISO as in Fig.~\ref{fig:bliss} gives a moderate bit error performance improvement over a channel SISO that outputs soft differential bits and an RLL SISO that with half the number of states. Even better bit error performance is possible by combining the channel SISO and the RLL SISO in one \textit{joint SISO} with the product space of the state space of the channel SISO and RLL SISO.}. 

Bliss schemes have pros and cons in comparison with a standard concatenation scheme:
\begin{itemize}
\item \textit{Pro:} In the Bliss scheme, the bit error rate (BER) at the input of the LDPC decoder, at least for the dominating systematic part, is the bare BER after bit-detection, which has not been multiplied by the error propagation factor of the RLL SISO decoder.
\item \textit{Con:} The error correction capability of the LDPC code is characterized by its code rate $\RLDPC$ (and secondarily by its block length $N$). The RLL encoding of the systematic part increases the number of bits into the ECC parity generator. For instance, for the same LDPC code rate $\RLDPC$ with a rate 2/3 RLL code, as applies for the $d=1$ case, the parity generator produces 1.5 times more parity bits, due to the RLL encoding of the systematic bit sequence. This additional amount of parity lowers the net rate of the Bliss scheme. 
\end{itemize}

In order to mitigate the rate loss of Bliss schemes due to the parity generation over redundant data as mentioned above, Immink \cite{Immink2} introduced a compression codec in Bliss' scheme. For storage channels with jitter noise on the transitions between run lengths, Zhang et al. \cite{Zhang} reduced the error propagation of the compression decoder for a $d=5$ RLL code by the use of Gray labeling. In this contribution, we take a different approach. We accept the redundancy of the systematic channel bit sequence, and modify the LDPC decoder to exploit this redundancy during its decoding iterations, in order to improve its error correction performance.  

Our interest is in the use of the same RLL codes by both modulation encoders and \textit{low density parity check} (LDPC) codes \cite{Gallager} as ECC codes with a high code rate (e.g. $\RLDPC=0.90-0.95$). Especially, in near-field optical storage as envisaged for the fourth generation of optical recording, RLL constraints with a minimum run length constraint of $d=1$ are still quite popular as evidenced by the recent RLL code design of Coene et al. \cite{Coene}. In \cite{Stitching}, we discussed a \textit{stitching technique} to connect a systematic part of a codeword to its adjacent parity part.

\section{Min-Sum LDPC Decoding Revisited}

\label{Section:MinSum}

The \textit{min-sum} LDPC decoding algorithm \cite{Fossorier} is a simplification of the sum-product algorithm that uses \textit{minimum and summation} operations instead of multiplication and summation operations. Both algorithms are a special case of the message passing algorithm \cite{Kschischang} that passes messages along the edges of a so called factor graph. These messages convey soft-decision information, e.g. in the form of \textit{log-likelihood ratios} (LLR). The min-sum algorithm always uses a \textit{log-likelihood} representation of the messages. For a binary random variable with probability $p_0$ ($p_1$) of taking on the value 0 (1), the log-likelihood equals $\log (p_0/p_1)$. With a properly chosen scaling factor after the minimum operation, the performance loss of the min-sum algorithm w.r.t. the sum-product algorithm is minimal \cite{Fossorier2} \cite{Heo}. The choice of the \textit{codeword bit node} or \textit{``variable node''} degrees in the \textit{factor graph} \cite{Kschischang} of the LDPC code seems to have only a minor influence for high LDPC code rate.

Let $N$ be the LDPC\footnote{Our method may also apply to other, similar codes with a low density parity check matrix, such as repeat-accumulate (RA) codes.} codeword length and $M$ be the number of parity check equations. The parity check matrix $H$ consists of $M$ rows and $N$ columns with elements from the binary Galois field $GF(2)$. For a parity check equation with index $m, \, 0 \leq m < M$, define the set $\mN(m)$ of codeword
bit positions that it checks, i.e.
\[
    \mN(m) = \{n \, | \, n=0,1,\ldots,N-1, \,  H_{mn} \neq 0 ).
\]
Similarly, for a codeword bit position $n, \, 0 \leq n < N$, define the set $\mM(n)$ of indices of parity check equations that check the given bit position $n$, i.e. 
\[
    \mM(n) = \{m \, | \, m=0,1,\ldots,M-1, \,  H_{mn} \neq 0 ).
\]
The \textit{factor graph} associated with the parity check matrix $H$ has as set of vertices $V$, the union of the set of $N$ bit nodes $V_{var}$ and a set of $M$ parity check nodes $V_{check}$. The set of edges $E$ consists of all edges $(m,n)$ for which $H_{mn} \neq 0$. For the sake of ease of discussion, from hereon, we assume that the LDPC code is \textit{regular}, which means that all sets $\mN(m)$ have the same size $K$, and that all sets $\mM(n)$ have the same size $J$. 

In general, the maximum likelihood or \textit{hard-decision} estimate of a binary random variable is determined by the sign of its log-likelihood value. The absolute value of a log-likelihood value is a \textit{reliability} of the \textit{hard-decision estimate}. A log-likelihood value of zero corresponds to an \textit{erasure} (no information). 

We now discuss the various operators involved in the min-sum algorithm. The operator following $\VAR$ that is applied inside the codeword bit nodes or ``variable nodes'' combines a number of sources of information (log-likelihood messages) about the bit associated with a variable node. These sources of information are assumed to be statistically independent. This assumption of \textit{statistical independence}\footnote{Due to the presence of cycles in the factor graph, this is only an approximation.} translates into a \textit{sum} operator on the input log-likelihoods $\Lambda_i, \ i=0,1,\ldots,J-1$ of a variable node.
\[
	\VAR(\Lambda_0,\Lambda_1,\ldots,\Lambda_{J-1}) = \Lambda_0 + \Lambda_1 + \ldots + \Lambda_{J-1}.\\    
\]
The operator $\CHK$ that is used inside the check nodes approximates the  log-likelihood value of the exclusive-or ``$\oplus$'' of its presumably statistically independent input variables. 
\begin{equation}
\begin{array}{l}
    |\CHK(\Lambda_0,\Lambda_1,\ldots,\Lambda_{K-1})| \\
    \ =   \alpha \min ( | \Lambda_0|, \, | \Lambda_1|, \ldots, |\Lambda_{K-1}| ) \\
\end{array}
\end{equation}
\begin{equation}
\label{eq:check_node}
\begin{array}{l}
    \sign(\CHK(\Lambda_0,\Lambda_1, \ldots, \Lambda_{K-1}))\\
    \ = \sign(\Lambda_0) \oplus \sign(\Lambda_1) \oplus 
    \ldots \oplus  \sign (\Lambda_{K-1}) \\
\end{array}
\end{equation}
where $0 < \alpha \leq 1$ is the aforementioned scaling factor and
\[
    \sign(x) =  \left\{ \begin{array}{ll}
                    0       & \mbox{if} \ x \geq 0, \\
                    1       & \mbox{if} \ x < 0. \\
                \end{array} \right.
\]

For a given iteration of the min-sum algorithm, we define the following variables.
As usual, in the message passing algorithm, messages are sent along edges of the factor graph.

\begin{itemize}
\item  $t_{n}$ -- \textit{The decoder input message into variable node
$n$}.

\item  $u_{mn}$ -- \textit{The message sent from variable node $n$ to check node $m$.} It is obtained as a function $\VAR$ of message $t_n$ and the last received messages $v_{mn}$ of
all check nodes $m', \, m' \neq m$,
\begin{equation}
\label{Eq:BitNodeOperation}
    u_{mn} = \VAR(  \, t_n, \, ( v_{m'n} \, | \, m' \in \mM(n) \backslash \{m\}) \, ). \\
\end{equation}

\item  $v_{mn}$ -- \textit{The message sent from check node $m$ to variable node $n$.} It is obtained as a function $\CHK$ of the last received messages $u_{mn}$ of all variable nodes $n', \, n' \neq n$,
\begin{equation}
\label{Eq:CheckNodeOperation}
    v_{mn} = \CHK(  \, ( u_{mn'} \, | \, n' \in \mN(m) \backslash \{n\}) \, ).
\end{equation}

\item  $w_{n}$ -- \textit{The decoder output messages.} Unlike the messages $u_{mn}$ the decoder output message $w_n$ uses all available information in a variable node $n$. It is obtained as the function $\VAR$ of the message $t_n$ and the last received messages $v_{mn}$ of \textit{all} check nodes $m$,
\begin{equation}
\label{Eq:OutputMessage}
    w_n = \VAR(  \, t_n, \, ( v_{m'n} \, | \,  m' \in \mM(n) ) \, ).
\end{equation}

\end{itemize}

A classical implementation of the min-sum algorithm stores all received messages. During the first half-iteration, all messages $u_{mn}$ are sent from all variable nodes to the check nodes. During the second half-iteration, all messages $v_{mn}$ are sent from all check nodes to the variable nodes. A constant number of iterations can be used. The decoder output messages need not be evaluated for all iterations, but only for the final iteration.

\section{$d=1$-Constraint Nodes in the Factor Graph}

Our approach is to modify the LDPC decoder so that it can use the knowledge that in the systematic channel sequence the subsequences $01^p0$ and $10^p1, \ p=1,2,\ldots,d$ do not occur. Here, $a^d$ denotes a sequence consisting of $d$ copies of the value $a, \ a=0,1$. This is achieved by addition of so called \textit{$d$-constraint nodes} in the LDPC code's factor graph. As stated before, our aim is to improve the error correction capability\footnote{Note, that a $d=1$ constraint itself does not imply a non-trivial Hamming distance $D, \ D > 1$. For instance the $d=1$ constrained sequences $00 \ldots 01100 \ldots 00$ and $00 \ldots 01110 \ldots 00$ evidently have Hamming distance $D=1$. However, for the kind of intermediate range of target bit error rates after LDPC decoding in our storage application\footnote{We assume that e.g. an outer RS (or BCH) codes is used to achieve the ultra-low bit error rates typical of storage applications.} a high minimum distance of the LDPC code is not required for good or even superior error correction performance, anyway.}  of the LDPC decoder. 

In the definition of a new, modified factor graph for the LDPC decoder in a Bliss scheme, it is very convenient that the $d$-constraint is a \textit{local} constraint \cite{Kschischang}. That means, that the constraint involves only a fixed, small number $d+2$ of codeword bits (codeword bit nodes). This sparseness of the connectivity matrix between the nodes is essential for efficient and effective operation of the modified LDPC decoder.  This is all the more true, the smaller the value of $d, \ d>0$ is. For this reason, and for the practical importance of their applications \cite{Coene}, from hereon we concentrate on the $d=1$-based Bliss schemes.  

\begin{figure}[H]
\centering
 \includegraphics[viewport = 50 420 400 710, width= 0.95\columnwidth, clip]{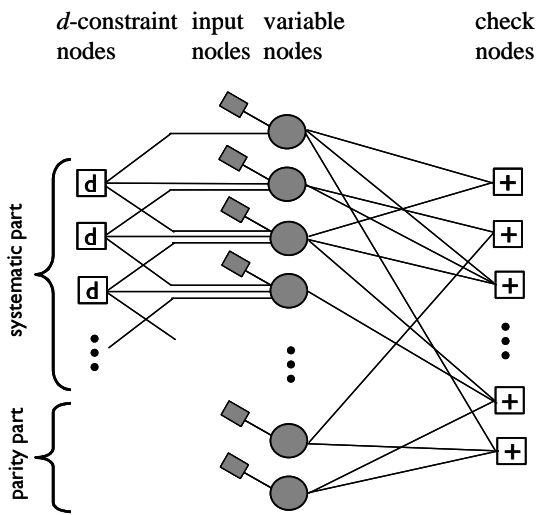}\\
 \caption{Factor graph of an LDPC code with additional $d$-constraint nodes for the case of $d=1$. 
  }\label{fig:factor_graph2}
 \end{figure}

The new factor graph with the $d$-constraint nodes is shown in Fig.~\ref{fig:factor_graph2}. In general, the degree of a $d$-constraint node equals $d+2$, as these nodes need to be able to detect the presence of the subsequences $01^d0$ and $10^d1$ of length $d+2$. Define the following additional variables in the LDPC decoding algorithm. 
\begin{itemize}
\item $a_{np}$ -- \textit{The input message to  $d$-constraint node $p$ sent from variable node $n$. }
\item $b_{np}$ -- \textit{The output message of $d$-constraint node $p$ sent to variable node $n$. }
\end{itemize}
Here, the index $n$ of the variable node (LDPC codeword variable node) is implicitly restricted to the systematic bit positions as there is no $d$-constraint on the parity part of an LDPC codeword. 

As a general principle in message passing algorithms, the output message along a certain arc from a certain node is only allowed to depend on the (most recently) received input messages via all \textit{other} arcs into that node . Hence, for  $d=1$-constraint node with index $n$, and most recently received log-likelihood messages $a_{n-1,n}, \ a_{n,n}$ and $a_{n+1,n}$, we need to specify three output functions $\CO_0, \CO_1$ and $\CO_2$ in order to generate as many log-likelihood output messages $b_{n-1,n}, \ b_{n,n}, \ b_{n+1,n}$, such that 
\[
\begin{array}{l}
	b_{n-1,n} = \CO_0 (a_{n,n}, a_{n+1,n})\\
	b_{n,n} = \CO_1 (a_{n-1,n}, a_{n+1,n})\\
	b_{n+1,n} = \CO_2 (a_{n-1,n}, a_{n,n}).\\	
\end{array}	
\]

Observe that, if one of the input log-likelihood values $ a_{n-1}, \ a_n, \ a_{n+1} $ of a $d=1$-constraint node is zero, there is no indication that the $d=1$-constraint is violated, as it is not possible to conclude to a violation from the knowledge of fewer than $d+2=3$ hard-decision estimates. Then, the output log-likelihood messages are to be zero. The output messages do not need to exert a force on the solution of the LDPC decoder in that case. On the contrary, if the input log-likelihoods have large absolute values, and their hard-decision values indicate a violation of the $d$-constraint, the output values, also should have large absolute values. However, if the hard-decision values indicate compliance with the $d$-constraint, the output log-likelihoods should be zero. Hence, in the spirit of the min-sum algorithm we choose to let
\[
	|\CO_r (x, \, y)| \in \{0, \ \min (|x|,\,|y| ) \, \}, \ \ r=0,1,2. 
\]
Similar to Eq. (\ref{eq:check_node}), one can use a scaling factor to post-multiply the above minimum values. 

Also the \textit{sign of} $\CO_0$ is not allowed to depend on $a_{n-1,n}$, etc. The signs are chosen such that they enforce the disappearance of violations of the $d=1$ constraint in the decoded LDPC codeword, see Tables \ref{tab:CO_0}, \ref{tab:CO_1} and \ref{tab:CO_2}. As stated before, a zero entry in these tables applies when there is no violation of the $d$-constraint.

\section{Min-Sum LDPC Decoding with $d=1$-Constraint Nodes in the Factor Graph}

With reference to the above definitions of log-likelihood messages, one iteration of the extension of the min-sum LDPC decoding algorithm with $d=1$-constraint nodes in the factor graph is given by the following equations. 

\begin{equation}
\begin{array}{llll} 
    u_{mn} & = \VAR( & t_n, \, ( v_{m'n} \, | \, m' \in \mM(n) \backslash \{m\} ), \\ 
    & &  \, b_{n-1,n}, \, b_{n,n}, \, b_{n+1,n} &) \\
    a_{n-1,n} &= \VAR( & t_n, \, ( v_{m'n} \, | \, m' \in \mM(n) ), \\ 
    	&  &b_{n,n}, \, b_{n+1,n} &) \\
    a_{n,n} &= \VAR( & t_n, \, ( v_{m'n} \, | \, m' \in \mM(n) ), \\  
    	& & b_{n-1,n}, \, b_{n+1,n} &) \\
    a_{n+1,n} & = \VAR( & t_n, \, ( v_{m'n} \, | \, m' \in \mM(n) ), \\  
    	&  & b_{n-1,n}, \, b_{n,n} & ) \\
    b_{n-1,n} &= \CO_0 (& a_{n,n}, \, a_{n+1,n} & ) \\
    b_{n,n} &= \CO_1 ( & a_{n-1,n}, \, a_{n+1,n} & ) \\
    b_{n+1,n} &= \CO_2 ( & a_{n-1,n}, \, a_{n,n} & ) \\
    v_{mn} & = \CHK( & ( u_{mn'} \, | \, n' \in \mN(m) \backslash \{n\}) & ) \\
    w_n & = \VAR(  &t_n, \, ( v_{m'n} \, | \,  m' \in \mM(n) ) & ) \\
\end{array}
\end{equation}

\section{Simulation Results}

\begin{figure}[H]
\centering
\includegraphics[viewport=60 270 510 630, width=0.95\columnwidth, clip]{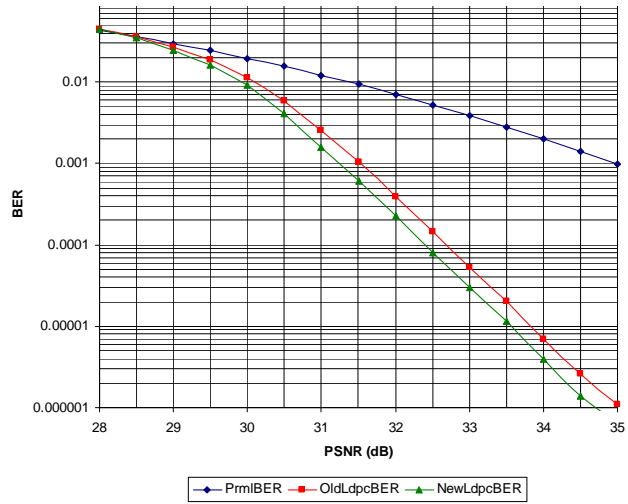}
\caption{Simulated BER curves for an LDPC-based Bliss scheme with the $J=3$-regular LDPC code is $\RLDPC=0.906$ and the code length $N=1728$. 
\label{fig:10b_fix_rawgb}}
\end{figure}

\begin{figure}[H]
\centering
\includegraphics[viewport=60 370 510 740, width=0.95\columnwidth, clip]{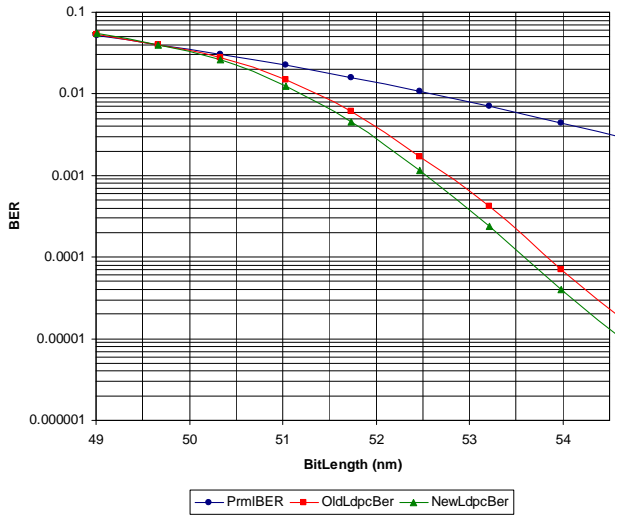}
\caption{Similar graph as Fig. \ref{fig:10b_fix_rawgb}, where the peak signal-to-noise ratio (PSNR) is fixed at 35 dB, and the bit-length of the channel bits is varied. 
\label{fig:10b_fix_psnr}}
\end{figure}

We experimented with two pseudo-random regular LDPC codes. The codeword bit nodes or variable bit nodes all had degree $J=3$. The shorter LDPC code had a code rate of $\RLDPC=0.906$ and a code length of $N=1728$. The bit error rates after LDPC decoding obtained from simulations using this shorter code are shown in Fig. \ref{fig:10b_fix_rawgb} and \ref{fig:10b_fix_psnr}. In Fig. \ref{fig:10b_fix_rawgb}, the peak signal-to-noise ratio (PSNR) of white additive noise at the channel output is varied for a fixed channel bit length of 53 nm, scaled to the numerical aperture and wavelength of Blu-ray disc \cite{Narahara}. The user capacity of the disc scaled to the physics of Blu-ray disc then equals 30.5 GB. The simulations used a channel model of the optical storage readout channel using the channel modulation transfer function from the Braat-Hopkins formalism  \cite{Braat}. The channel bit length is scaled to the physical readout parameters of Blu-ray disc. In Fig. \ref{fig:10b_fix_psnr}, channel bit length is varied, at a fixed PSNR of 35 dB. The $d=1$ RLL code of \cite{Coene} is used, that has code rate $\RRLL=2/3$. 

The number of LDPC iterations was set at 16. We use the schedule of Yeo et al. \cite{Yeo}, wherein a \textit{single} check node or $d$-constraint node update is followed by variable updates of the connected variable nodes as a kind of ``mini decoding iterations.'' We process the check nodes and the $d$-constraint nodes in sequence. An LDPC decoding iteration is then complete when all check nodes and all $d$-constraint nodes have undergone an update. This update schedule about halves the number of iterations that need to be performed in a decoder. Up to half a million LDPC codeword have been simulated per point of the graph. 

In all graphs presented in this section, the top curve shows the BER after bit-detection. The middle curve shows the BER after standard LDPC decoding. The bottom curve shows the BER with decoding using the additional RLL-constraint nodes in the factor graph. 

\begin{figure}[H]
\centering
\includegraphics[viewport=60 270 510 630, width=0.95\columnwidth, clip]{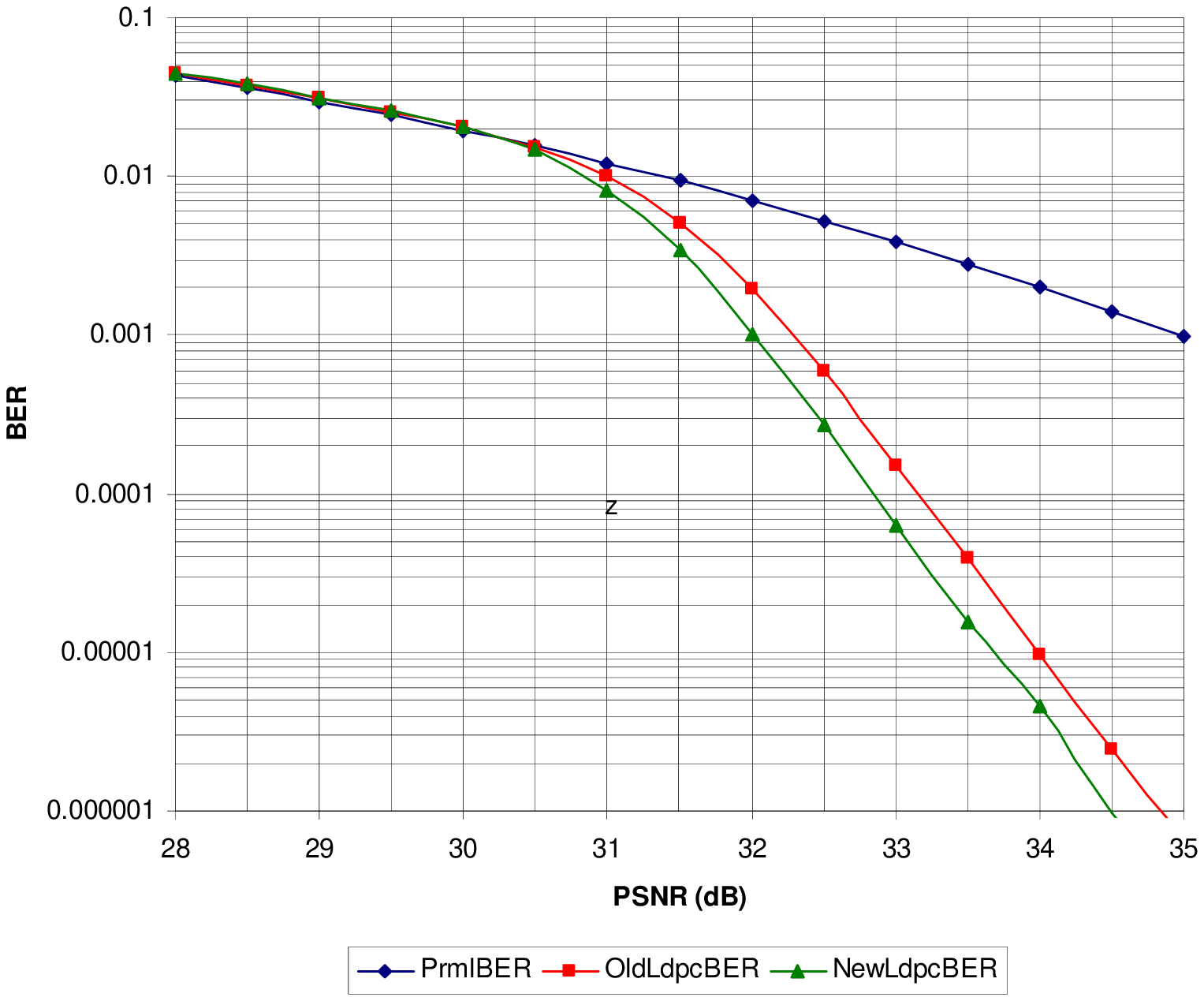}
\caption{Similar graph as Fig. \ref{fig:10b_fix_rawgb}, with a longer LDPC code (length 6912 bits) of a higher code rate 0.955. 
\label{fig:8b_fix_rawgb}}
\end{figure}

\begin{figure}[H]
\centering
\includegraphics[viewport=60 270 510 630, width=0.95\columnwidth, clip]{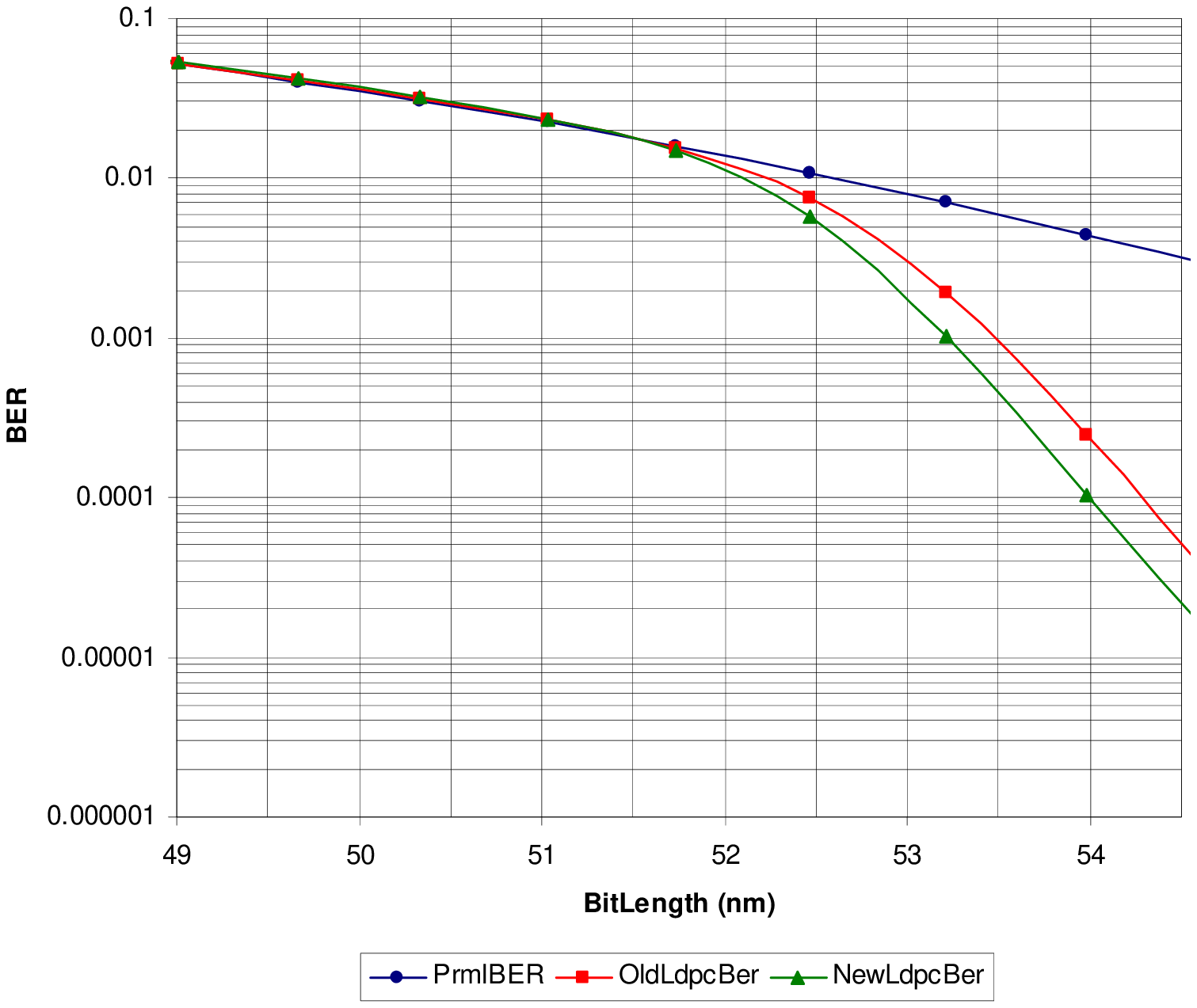}
\caption{Similar graph as Fig. \ref{fig:10b_fix_psnr}, also with a longer LDPC code (length 6912 bits) of a higher code rate 0.955, as in Fig. \ref{fig:8b_fix_rawgb}. 
\label{fig:8b_fix_psnr}}
\end{figure}

The longer LDPC code has a code rate of $\RLDPC=0.955$ and a code length of $N=6912$. The BER simulation results for this longer code are shown in Fig. \ref{fig:8b_fix_rawgb} and \ref{fig:8b_fix_psnr}. 

For a physical sector size of the storage medium of approx. 1 Mbit, the shorter LDPC code length allows combination with an \textit{10-bit outer RS code} with about maximal length ($2^{10}-1$ symbols). The longer LDPC code allows combination with an \textit{8-bit outer RS code} of about maximal length ($2^8-1)$.

\section{Conclusion}

We conclude that the use of a $d=1$-constraint during the decoding of an LDPC-based Bliss scheme, brings an advantage in PSNR of around 0.25 dB. Using this technique, the bit length of a simulated model of an optical storage channel can be decreased by ca. 0.6 \%. This suggests that the use of $d=1$-constraint during LDPC decoding can enable an increase in the storage density of around 0.6 \%.


%




\section*{Acknowledgment}

These results were obtained during a joint project between Philips Research and Sony. 
The author is indebted to his former colleague Haibin Zhang, now with TNO Telecom, for the construction of the LDPC codes used in this research. 
Furthermore, the author would like to acknowledge stimulating discussions with his former colleagues Stan Baggen, Wim Coene and Bin Yin as well as his project members from Sony, Seji Kobajashi, Toshi Horigome, Makato Noda and Hiroyuki Yamagischi.





\begin{thebibliography}{99}
\bibitem{Bliss} W. G. Bliss,
``Circuitry for performing error correction calculations on
baseband encoded data to eliminate error propagation,'' \emph{IBM Techn. Discl. Bul.}, vol. 23, pp. 4633-4634, 1981.
\bibitem{Fan} J. L. Fan, \emph{Constrained Coding
and Soft Iterative Decoding}, Norwell, MA: Kluwer Academic
Publishers, 2001, pp. 144-145.
\bibitem{Immink} K. A. S. Immink, \emph{Codes for mass data
storage systems}, 2nd ed., Eindhoven, The Netherlands: Shannon
Foundation Publishers, 2004, pp. 51-64.
\bibitem{Gallager} R. G. Gallager, \emph{Low-Density
Parity-Check Codes}, Cambridge, MA: MIT Press, 1963.
\bibitem{Coene} W. M. J. Coene, A. Hekstra, B.
Yin, H. Yamagishi, M. Noda, A. Nakaoki, and T. Horigome, ``A new d=1, k=10 soft-decodable RLL code with r=2 RMTR-constraint and a 2-to-3 PCWA mapping for DC-control,'' in \emph{Proc. of SPIE}, Vol 6282,  Optical Data Storage 2006, R. Katayoma, R. Schlesinga, editors.
\bibitem{Stitching} A.P. Hekstra, W.M.J. Coene, R.J.W Debets, ``A Stitching Technique for Bliss Schemes,'' submitted to \emph{IEEE Tr. on Magn.}, May 2007. 
\bibitem{Immink2} K. A. S. Immink, ``A practical method for
approaching the channel capacity of constrained channels,''
\emph{IEEE Trans. Inform. Theory}, vol. 43, no. 5, pp. 1389-1399,
Sept. 1997. 
\bibitem{Zhang} H. Zhang, A. P. Hekstra, W.M.J. Coene and B. Yin, ``Performance investigation of soft-decodable RLL codes in high density optical recording,'' accepted for publication in \emph{IEEE Trans. on Magn.}, 2007.
pp. 879-887, Aug. 2006.
\bibitem{Fossorier} M. Fossorier, M. Mihaljevic, H. Imai, ``Reduced complexity
iterative decoding of low-density parity check codes based on belief
propagation,'' \emph{IEEE Trans. on Commun.}, Vol. COM-47, No. 5,
pp. 673-680, May 1999.

\bibitem{Kschischang} F.R. Kschischang, B. J. Frey, H.A. Loeliger,
``Factor Graphs and the Sum-Product
Algorithm,'' \emph{IEEE Trans. Inf. Th.}, Vol. IT-47, No. 2, Febr.
2001.

\bibitem{Fossorier2} J. Chen, M. Fossorier, ``Near optimum universal belief
propagation based decoding of low density parity check codes,''
\emph{IEEE Trans. on Commun.}, Vol. COM-50, No. 3, pp. 406-414,
March 2002.

\bibitem{Heo}J. Heo, ``Analysis of scaling soft information on low density
parity check code,'' \emph{Electron Lett.}, vol. 39, no. 2, pp.
219-221, Jan. 2003.

\bibitem{Narahara} T. Narahara, S. Kobayashi, M. Hattori, Y. Shimpuku, G. van den
Enden, J. Kahlman, M. van Dijk, R. van Woudenberg, ``Optical disc
system for digital video recording,'' \emph{Proc. ISOM/ODS 1999},
Hawaii, Jul. 1999, SPIE Vol. 3864, pp. 50-52.

\bibitem{Braat} J. Braat, \emph{Principles of Optical Disc Systems: Chapter2, Read-out of Optical Discs,} Adam Hilger, Ltd, 1985.  

\bibitem{Yeo} E. Yeo, B. Nikolic, V. Anantharam, ``Architectures and implementations of low-density parity check decoding algorithms,'' \emph{The 2002 45th Midwest Symp. on Circuits and Systems}, Vol. 3, Aug. 2002, pp. III-437-III-440. 

\end{thebibliography}
%

\begin{figure}[H]
\centering
 \includegraphics[viewport = 50 420 340 710, width= 0.8\columnwidth, clip]{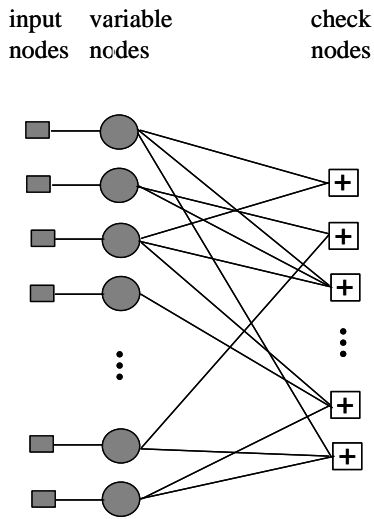}\\
 \caption{Factor graph of an LDPC code. Variable nodes are located in the middle. At the left are the input nodes, representing the channel output messages. Check nodes are situated at the right. 
  }\label{fig:factor_graph}
 \end{figure}

\begin{table}[H]
\centering
\begin{tabular}{cc|c}
$a_{n-1,n}$ & $a_{n+1,n}$ & $b_{n,n}=CO_1 (a_{n-1,n}, a_{n+1,n})$ \\
\hline
\hline
+ & + & + \\
+ & - & 0 \\
- & + & 0 \\
- & - & - \\
\end{tabular}
\caption{Sign of the output log-likelihood message $b_{n,n}$ of the $d$-constraint node with index $i$ as determined by the function $CO_1 (a_{n-1,n}, a_{n+1,n})$ }
\label{tab:CO_0}
\end{table}

\begin{table}[H]
\centering
\begin{tabular}{cc|c}
$a_{n-1,n}$ & $a_{n,n}$ & $b_{n+1,n}=CO_2 (a_{n-1,n}, a_{n,n})$ \\
\hline
\hline
+ & + & 0 \\
+ & - & - \\
- & + & + \\
- & - & 0 \\
\end{tabular}
\caption{Sign of the output log-likelihood message $b_{n+1,n}$ of the $d$-constraint node with index $i$ as determined by the function $CO_2 (a_{n-1,n}, a_{n,n})$ }
\label{tab:CO_1}
\end{table}

\begin{table}[H]
\centering
\begin{tabular}{cc|c}
$a_{n,n}$ & $a_{n+1,n}$ & $b_{n-1,n}=CO_0 (a_{n,n}, a_{n+1,n})$ \\
\hline
\hline
+ & + & 0 \\
+ & - & - \\
- & + & + \\
- & - & 0 \\
\end{tabular}
\caption{Sign of the output log-likelihood message $b_{n-1,n}$ of the $d$-constraint node with index $n$ as determined by the function $CO_2 (a_{n,n}, a_{n+1,n})$ }
\label{tab:CO_2}
\end{table}

%

\begin{IEEEbiography}[{\includegraphics[width=1in,height=1.25in,clip,keepaspectratio]{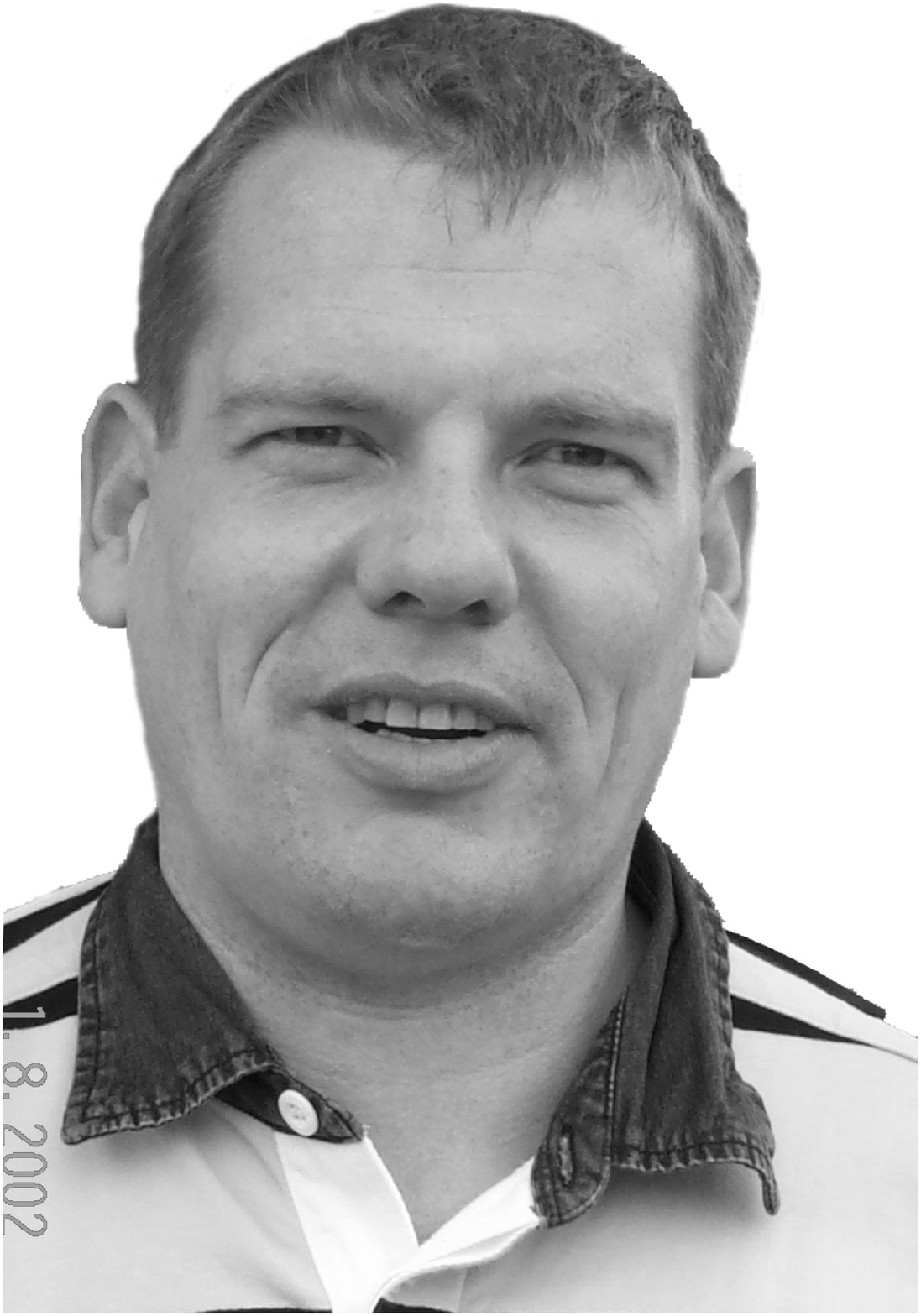}}]{Andries P. Hekstra}
(M'00) was born in Breda, the Netherlands in 1961. He received the "Ingenieur" degree in Electrical Engineering from Eindhoven University, the Netherlands, summa cum laude in 1985 with a specialization in multi-user information theory. In 1985-86 he was a Young Graduate Trainee at the European Space Agency where he worked on spread spectrum telemetry systems in Darmstadt, Germany. Successively, he was a Ph.D. student at the Electrical Engineering Department of Cornell University, Ithaca, USA. There, he studied abstractions of VLSI
packing problems. In 1990 he joined KPN Research in Leidschendam, the Netherlands, where he finished his Ph.D. degree in 1994 at Eindhoven University of Technology with his professor from Cornell University as second promoter. From 1995 to 2000 he investigated automatic assessment of video and speech quality using models of human perception and cognition. During 2001-2006, he worked at Philips Research, Eindhoven, mainly on error correction for optical storage systems. Since Sept. 2006, he works on applied communication and information theory at large within the research department of NXP Semiconductors, Eindhoven. 
\end{IEEEbiography}







\end{document}